\documentclass[aps,superscriptaddress,footinbib,twocolumn, amsmath,amssymb,
 aps]{revtex4-1}
\usepackage{graphics}      
\usepackage{graphicx}      
\usepackage{longtable}     
\usepackage{url}           
\usepackage{bm}            
\usepackage[usenames]{color}
\usepackage[dvipsnames]{xcolor}
\usepackage{cancel}
\usepackage{soul}
\usepackage{ulem}
\usepackage{amsmath,amssymb}
\usepackage{epstopdf}

\begin{document}

\preprint{APS/123-QED}

\title{Expected and unexpected routes to synchronization in a system of swarmalators}

\author{Steve J. Kongni}
\affiliation{Research Unit Condensed Matter, Electronics and Signal Processing, University of Dschang, P.O. Box 67 Dschang, Cameroon.}
\affiliation{MoCLiS Research Group, Dschang, Cameroon.}

\author{Thierry Njougouo}

\affiliation{IMT School for Advanced Studies, Lucca, Italy}
\affiliation{Faculty of Computer Science and Namur Institute for Complex Systems (naXys), University of Namur, Namur, Belgium}
\affiliation{Department of Electrical and Electronic Engineering, Faculty of Engineering and Technology (FET), University of Buea, Buea, Cameroon}
\affiliation{ MoCLiS Research Group, Dschang, Cameroon.}

\author{Patrick Louodop}
\affiliation{Research Unit Condensed Matter, Electronics and Signal Processing, University of Dschang, P.O. Box 67 Dschang, Cameroon.}
\affiliation{ICTP South American Institute for Fundamental Research,
S\~ao Paulo State University (UNESP), Instituto de F\'{i}sica Te\'{o}rica,
Rua Dr. Bento Teobaldo Ferraz 271,
Bloco II, Barra Funda, 01140-070 S\~ao Paulo, Brazil.}
\affiliation{MoCLiS Research Group, Dschang, Cameroon.}

\author{Robert Tchitnga}

\affiliation{Research Unit Condensed Matter, Electronics and Signal Processing, University of Dschang, P.O. Box 67 Dschang, Cameroon.}

\author{Fernando F. Ferreira}
\affiliation{Department of Physics-FFCLRP, University of S\~ao Paulo, Ribeirao Preto-SP, 14040-901, Brasil.}

\author{Hilda A. Cerdeira}

\affiliation{S\~ao Paulo State University (UNESP), Instituto de F\'{i}sica Te\'{o}rica, Rua Dr. Bento Teobaldo Ferraz 271, 
Bloco II, Barra Funda, 01140-070 S\~ao Paulo, Brazil.\\
Epistemic, Gomez $\&$ Gomez Ltda. ME, Rua Paulo Franco 520, Vila Leopoldina, 05305-031 S\~ao Paulo, Brazil\\}

\begin{abstract}
Systems of oscillators whose internal phases and spatial dynamics are coupled, swarmalators,  present diverse collective behaviors which in some cases lead to explosive synchronization in a finite population as a function of the coupling parameter between internal phases. Near the synchronization transition, the phase energy of the particles is represented by the XY model, and they undergo a transition which can be of the first order or second depending on the distribution of natural frequencies of their internal dynamics. The first order transition is obtained after an intermediate state (Static Wings Phase Wave state (SWPW)) from which the nodes, in cascade over time, achieve complete phase synchronization at a precise value of the coupling constant. For a particular case of natural frequencies distribution, a new phenomenon of Rotational Splintered Phase Wave state (RSpPW) is observed and leads progressively to synchronization through clusters switching alternatively from one to two and for which the frequency decreases as the phase coupling increases.

\end{abstract}

\maketitle

Studies on synchronization of dynamical systems started with Christian Huygens in 1665 \textcolor{blue}{\cite{huygens1895letters}}. This process describes the coherent behavior of interconnected dynamical systems and is commonly observed in research fields such as biology where we find it in the rhythmic evolution of a pacemaker with heartbeats \textcolor{blue}{\cite{giuriato2020timed,lyu2020synchronized}}, the flocking birds \textcolor{blue}{\cite{cavagna1, chate, bajek}}, flocking fish \textcolor{blue}{\cite{fish, bellomo2017quest, ballerini2008interaction}}. In physics, it is seen in many cases, such as in the aforementioned synchronous displacement of two pendulums or the synchronized state of two or more oscillators when an intermittent coupling is considered \textcolor{blue}{\cite{huygens1895letters, pec1990, sbocca2002, njougouo2020effects}}. 
\\
\\
\quad Building on the foundational work of Huygens, subsequent studies have expanded the understanding of synchronization across various domains. Notably, A.T. Winfree introduced a model in 1967 that employed coupled oscillators to examine circadian rhythms, marking a significant advancement in the study of biological synchronization \textcolor{blue}{\cite{winfree1967biological}}. After the fundamental work by Kuramoto, who introduced a simplified model to describe phase synchronization in systems with a large number of oscillators \textcolor{blue}{\cite{kuramoto2002coexistence}}, research has further expanded into the realm of complex networks. Synchronization phenomena in these networks have been extensively explored, including in multi-layer configurations, both with and without amplification effects \textcolor{blue}{\cite{njougouo2020dynamics, withou}}. Moreover, the practical applications of synchronization have been demonstrated across various contexts \textcolor{blue}{\cite{tang}}. The study of coupled chaotic oscillators has also contributed significantly to understanding synchronization dynamics \textcolor{blue}{\cite{louodop2017coherent, frasca2018, alex2008}}. Given the breadth of these investigations, the transition to synchronization in complex networks, particularly in systems with mobile components, has garnered substantial interest and has been the subject of extensive research \textcolor{blue}{\cite{arenas2008synchronization,nguefoue2021network,ji2014analysis}}.
\\
\\
\quad Recently, a model was introduced specifically for mobile systems, known as swarmalators \textcolor{blue}{\cite{o2017oscillators}}. This model couples the spatial positions of systems with the phase dynamics of the Kuramoto model. The authors showed that the swarmalators system presents five main patterns of behavior depending on the coupling strength between spatial and phase dynamics\textcolor{blue}{\cite{o2017oscillators,barcis2020sandsbots,o2018ring}}. Later, J. U. Lizarraga and M. A. Aguiar \textcolor{blue}{\cite{lizarraga2020synchronization}} improved the model proposed by K. P. O'Keeffe in \textcolor{blue}{\cite{o2017oscillators}} by considering the effect of an external force. They showed that the swarmalators system could synchronize or present aggregation patterns when the amplitude of the external force increases. 
\\
\\
Another particularly intriguing phenomenon obtained in complex networks is the explosive collective transition to synchronization. Gómez Gardeñes et al. demonstrated that explosive synchronization can occur in scale-free Kuramoto networks \textcolor{blue}{\cite{gomez2011explosive}}, while Skardal et al. \textcolor{blue}{\cite{skardal2014disorder}} revealed that the presence of disorder in the natural frequencies, depending on its amplitude, can also trigger explosive transitions. Similarly, other studies have uncovered explosive synchronization in both adaptive and multilayer networks \textcolor{blue}{\cite{zhang2015explosive,peron2012explosive,hong2022first}}. \\

Recent findings had also shown that the 2D swarmalators systems can exhibit first-order transition under the effect of attractive and repulsive interaction in a multiplex of swarmalators \textcolor{blue}{\cite{kongni2023phase}}. Based on the 2D model of swarmalators proposed in \textcolor{blue}{\cite{o2017oscillators}}, Sar et al. developed 1D model where they show that a random pinning on the swarmalators can lead a system to a chaotic behaviour \textcolor{blue}{\cite{sar2023pinning}}, while others investigated the existence of anti-phase synchronization between two groups of swarmalators \textcolor{blue}{\cite{kongni2023phase, ghosh2023antiphase}}. While the aforementioned studies and others \textcolor{blue}{\cite{hong2023swarmalators, lizarraga2023synchronization, sar2023swarmalators}} have explored the swarmalators model primarily through pairwise interactions, real-world systems often involve more complex dynamics. To bridge this gap, a new model has been proposed that incorporates higher-order interactions, considering relationships that extend beyond just two elements within a population \textcolor{blue}{\cite{anwar2023collective}}. This approach aims to better capture the complexity of interactions observed in real systems.\\
 
In this work we highlight the effect of the internal dynamics' phase coupling strength on the spatial dynamics of the swarmalators and vice versa. We investigate the influence that the natural frequencies of the swarmalators' internal dynamics have in the transition to phase synchronization and found that explosive or first order phase transition is not always a characteristic of these systems. \\ 

We consider a model of identical swarmalators confined to move in a two-dimensional space as defined by O'Keefe et al. \textcolor{blue}{\cite{o2017oscillators}} where the position and the phase of each entity are coupled and described by the following Eqs.\ref{e1} and \ref{e2}:

\begin{equation}\label{e1}
{{\dot X}_i}=v_i +\frac{1}{N}\sum\limits_{j \ne i}^N {\left[ {{F_{att}}\left( {{X_j}- {X_i}} \right)W\left( {{\theta _j} - {\theta _i}} \right)-{F_{rep}}\left( {{X_j} - {X_i}} \right)} \right]}, \\
\end{equation}

\begin{equation}\label{e2}
{{\dot \theta }_i} = w_i + \frac{K}{N}\sum\limits_{j \ne i}^N {{H_{att}}\left( {{\theta _j} - {\theta _i}} \right)G\left( {{X_j} - {X_i}} \right),} 
\end{equation}\\

with $i,j=1,2...N$ where $\theta_i$ is the phase of the internal dynamics of each swarmalator, represented by a Kuramoto-like model, ${X_i} = {\left( {{x_{i}},{y_{i}}} \right)^T}$ is the position coordinates in space of the $i^{th}$ entity, $N$ is the size of the population of swarmalators and $v_i$ and $w_i$ are  the velocity and natural frequency of each element respectively. We suppose that the dynamics of a system is defined by the spatial angle $\phi$, which describes its position defined by $\phi_i={\tan^{-1}}(y_i/x_i)$, and the coupled phase $\theta_i$. 

The model presents an attractive and a repulsive behavior due to the existence of two interaction forces: a long and a short range interaction, represented by the spatial interactions $F_{att}$ and $F_{rep}$ respectively and $H_{att}$ which is the phase interaction \textcolor{blue}{\cite{o2017oscillators,lizarraga2020synchronization}}. The competition between $F_{att}$ and $F_{rep}$ gives rise to clusters of particles with sharp boundaries, in agreement with many biological systems \textcolor{blue}{\cite{o2019review}.} The functions $W$ and $G$ represent the influence of the internal dynamics on the oscillators' movement and vice-versa, respectively.  Thus, the model presented previously in Eqs.\ref{e1} and \ref{e2} can be rewritten as follows:

\small{
\begin{equation}\label{ee1}
{{\dot X}_i} = v_i+\frac{1}{N}\sum\limits_{j \ne i}^N {\left[ {\frac{{{X_j} - {X_i}}}{{\left. {\left| {{X_j} - {X_i}} \right.} \right|}}\left( {A + J\cos \left( {{\theta _j} - {\theta _i}} \right)} \right)- B\frac{{{X_j} - {X_i}}}{{{{\left. {\left| {{X_j} - {X_i}} \right.} \right|}^2}}}} \right]},
\end{equation}}

\begin{equation}\label{ee2}
  {{\dot \theta }_i} = w_i+\frac{K}{N}\sum\limits_{j \ne i}^N {\frac{{\sin \left( {{\theta _j} - {\theta _i}} \right)}}{{\left. {\left| {{X_j} - {X_i}} \right.} \right|}}}.
\end{equation} 

\begin{figure*}[htp]
\includegraphics[width=18.5cm, height=8.5cm]{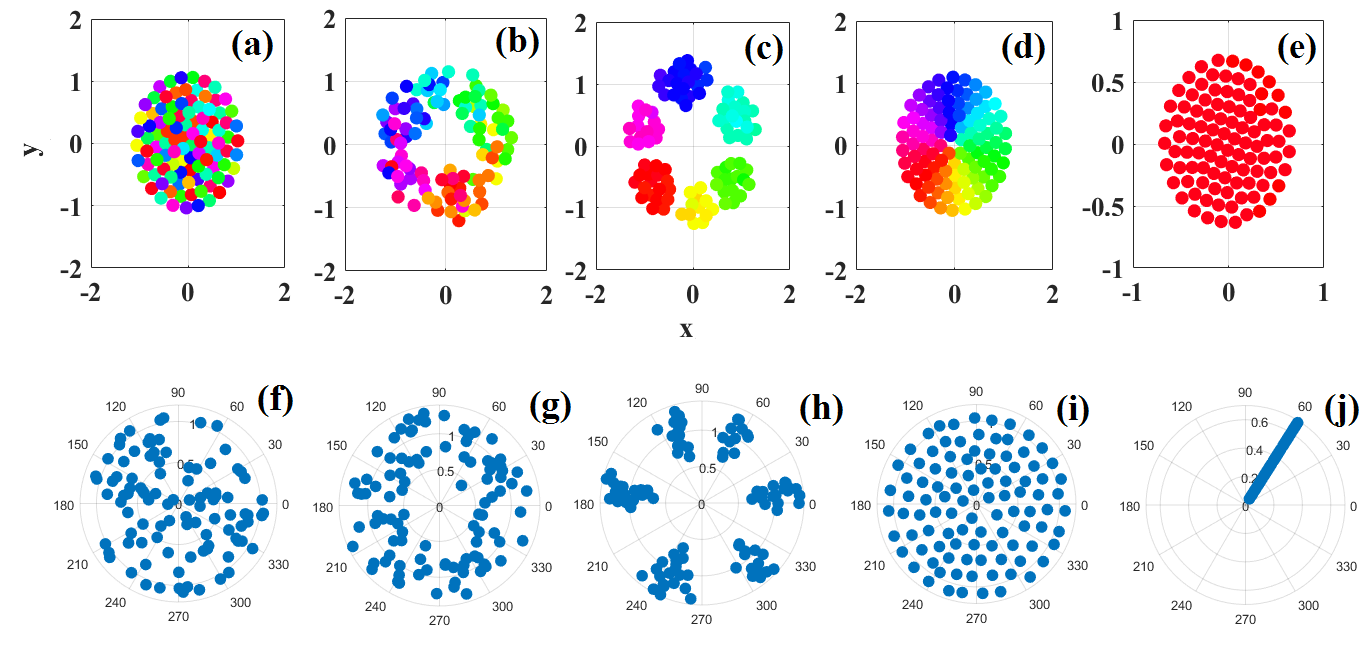}
\caption{Pattern formations of $N=100$ swarmalators presented in \textcolor{blue}{\cite{o2017oscillators, lee2021collective}} for different values of coupling parameters $J$ and $K$. Scatter plot of 
the network and corresponding internal phase states, showing: a) and f) Static Async state $(J=0.1, K=-1)$; b) and g) Active Phase Wave $(J=1,K=-0.75)$; c) and h) Splintered Phase Wave $(J=1,K=-0.1)$;  d) and i) Static Phase Wave state $(J=0.1,K=0)$; and e) and j) Static Sync state $(J=0.1,K=1)$. Notice that the name of the phases refers to the spatial distribution of the agents.}
\label{Other}
\end{figure*}

In Eq. \ref{ee1}, the interaction between the oscillators in space is modulated by the term $A + J\cos \left( {{\theta _j} - {\theta _i}} \right)$ with $A=B=1$. For simplicity, we choose $v_i=v$ and $\omega_i=\omega$, therefore, we take $v$ and $\omega$ equal to zero without loss of generality. K represents the phase coupling and J the attraction or repulsion between the particles of the system. Indeed, depending on the value of $J$ there can be attraction or repulsion between these elements. 
For a positive value of coupling $J$, we have an attraction between particles with the same phase. Conversely an opposite behavior is observed when $(J<0)$ \textcolor{blue}{\cite{o2017oscillators, lizarraga2020synchronization, jimenez2020oscillatory}}. Summarizing, depending on the values of the pair $(J,K)$, the dynamics of the swarmalators varies from aggregation to synchronization, and it can be seen in the Fig.\ref{Other}.  
\textcolor{blue}{\cite{o2017oscillators,barcis2020sandsbots, o2018ring, lizarraga2020synchronization,jimenez2020oscillatory, lee2021collective}}.\\

 The dynamics of swarmalators impose that they present only phase synchronization since two agents cannot occupy the same spatial position but can nevertheless have the same internal phase.
Per current literature, we use the order parameter $R$ defined by Kuramoto \textcolor{blue}{ \cite{moreira2019global, jeon2018recurrent, chen2017order,kuramoto2002coexistence}} to characterize synchronization

\begin{equation}\label{orderpar}
 R{e^{l\Phi }} = \frac{1}{N}\sum\limits_{i = 1}^N {{e^{l{\theta _i}}}}, \,\,\, with \,\,\,\,\ l^2=-1.
\end{equation}

When the internal dynamics of the $i^{th}$ and $j^{th}$ particles are synchronized: $\theta_i=\theta_j$ the norm $R$ tends to $1$, otherwise $R$ tends to $0$.
In the same vein, to measure the correlation between the spatial angle (angular position) $\phi$ and the internal phase dynamics $\theta$ of swarmalators we define another order parameter $S$ \textcolor{blue}{\cite{barcis2020sandsbots, o2017oscillators, lizarraga2020synchronization, jimenez2020oscillatory}}. 

\begin{equation}\label{ordercomp}
 {S_ \pm }{e^{l{\Psi _ \pm }}} = \frac{1}{N}\sum\limits_{i = 1}^N {{e^{l\left( {{\phi _i} \pm {\theta _i}} \right)}}},  \,\,\, with \,\,\,\,\ l^2=-1, 
\end{equation}

where $S=max(S_+,S_-))$ is the
real part of the complex order parameter, if $S=1$ there is full correlation between $\phi$ and $\theta$, if $S=0$ (or less than 1) it indicates lack of correlation.\\

\begin{figure}
\includegraphics[width=7.5cm, height=5cm]{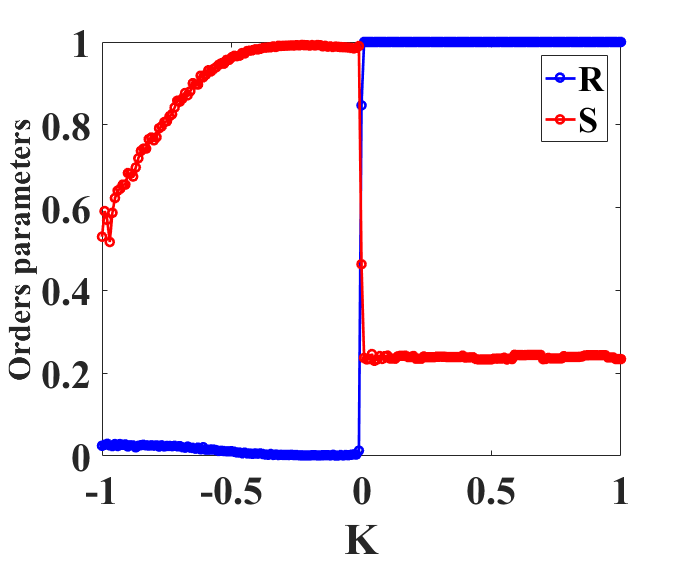}

\caption{Influence of internal dynamics' phase coupling K on the spatial dynamics of swarmalators represented by the Order Parameter, $R$, and Complex Order Parameter, $S$. Step of K equal to $0,01$ and spatial coupling $J=1$ (See movie 1 in supplemental material \textcolor{blue}{\cite{suppl}} where we show the evolution of the dynamics before and at the transition to synchronization as a function of the phase coupling $K$). }
\label{orders}
\end{figure}

We consider a system of $N = 50$ units in a two dimensional space. The initial conditions of the positions are uniformly and randomly selected between $[-1,1]$ with an initial phase $\theta$ between $[-\pi, \pi]$. To solve the set of differential Eqs.\ref{ee1} and \ref{ee2} we use the 4th order Runge Kutta integration algorithm with an integration step $dt=0.05$ and we study the long-term behavior reached by the system of swarmalators
after $20, 000$ iterations. \\
\\
\quad As defined previously, we use the Kuramoto order parameter to characterize the synchronization in the network. Fig.\ref{orders} presents its evolution as a function of the coupling parameter $K$ where we notice an explosive transition from no synchronization between swarmalators for $K<0$ to phase synchronization for $K>0$.\\
    
\begin{figure}
\includegraphics[width=7.5cm, height=5cm]{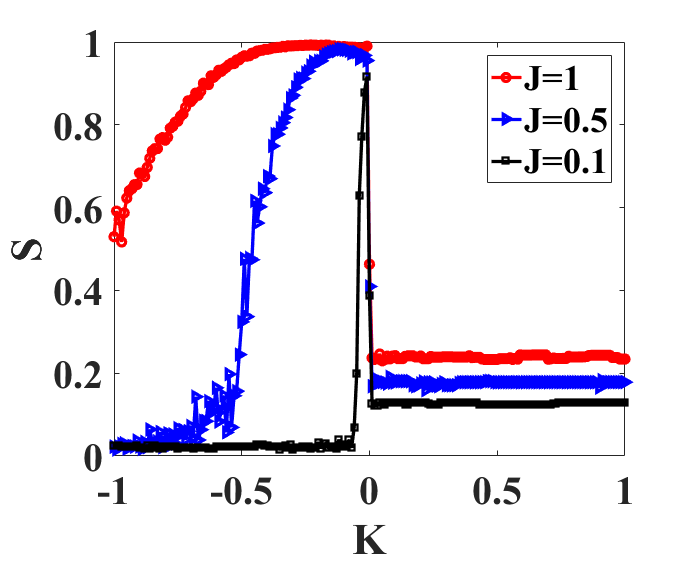}
\caption{Order parameter $S$ versus coupling $K$ showing the evolution of the correlation between $\theta$ and $\phi$ for different values of the phase coupling $J$.}.
 \label{orderS}
\end{figure}

Several authors have already shown that the explosive transition affects the correlation between the spatial and internal dynamics of the swarmalators, which depends on the coupling constant $K$ \textcolor{blue}{\cite{lizarraga2020synchronization,o2017oscillators,o2018ring,barcis2020sandsbots,hong2021coupling}}. 
\textcolor{blue}{In particular} this correlation happens on the SpPW state because the spatial dynamics is highly influenced by the phases that synchronize in clusters before complete synchronization occurs. In Figs. \ref{orders} and \ref{orderS}, it is also shown that, for $K > 0$,  the correlation between spatial and phase dynamics of the swarmalators is drastically reduced when the system synchronizes.
For the Splintered Phase Wave and Active Phase Wave states, the swarmalators move around space \textcolor{blue}{\cite{hong2018active,o2017oscillators,hong2021coupling}} and we calculate, that $S>0$, the mean velocity $V$ is non zero and positive $(V>0)$ (with $V = \frac{1}{N}\sum\limits_{i = 1}^N {\sqrt {\dot x_i^2 + \dot y_i^2} }$) and $R=0$. $S=1$, which implies that the phase and the spatial angle are perfectly correlated, is only possible on the Splintered and Static Phase Wave states shown in (Fig.\ref{Other}(c,h) and (d,i)). 

The phase coupling parameter's negative range is associated to the correlation range between phase and spatial dynamics ($S \approx 1$). Furthermore, the correlation $S$ is also influenced by the spatial coupling strength $J$. Indeed for $K<0$ and for small value of $J$ there is no correlation between $\theta$ and $\phi$ (this means $S \simeq 0$) and the correlation appears when $J$ increases (always for $K<0$), as can be seen in Fig.\ref{orderS}.  We observe no correlation for a synchronized state $(K>0)$.
 
\begin{figure}[ht]
\includegraphics[width=8.5cm, height=5cm]{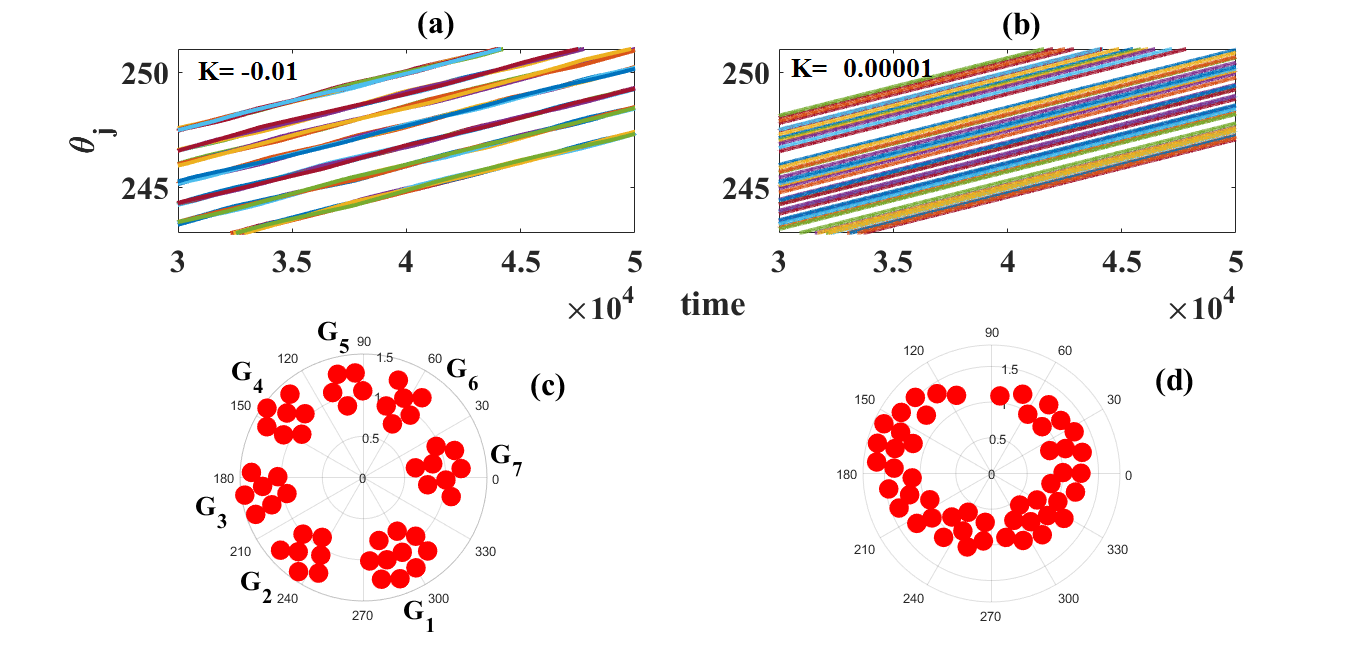}
\caption{Route to synchronization (the state SpPW, figures (a) and (c), is shown in movie 2 of the supplemental material \textcolor{blue}{\cite{suppl}}) where both the spatial and phase evolution are presented in a polar representation.}
 \label{rexp}
\end{figure}

\begin{figure}[ht]
\includegraphics[width=8cm, height=5.5cm]{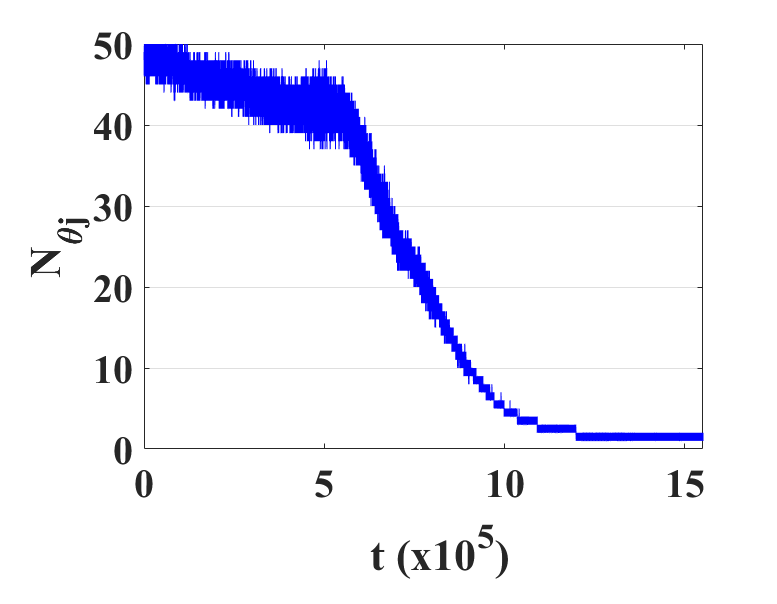}
\caption{Time dynamics of the swarmalators for $K=10^{-4}$ showing the change of the number of nodes that are out of synchronization. Here the system moves from SWPW to SS states through cascade - route - to synchronization at the critical value of $K$.}.
 \label{casc}
\end{figure}

To reach explosive synchronization for $J = 1$, the swarmalators move from Splintered Phase Wave state (SpPW), shown by the graphs on Figs.\ref{rexp} (a) and (c), to Static Wing Phase Wave state (SWPW) (See graphs on Figs.\ref{rexp} (b) and (d)) (with properties similar to Static Phase Wave State) before they get to the Static Sync state (SS). This new state is an unexpected path to synchronization. The stability of the new state SWPW is shown in the Appendix \ref{Appx}. 
The transition, from SWPW to SS, occurs at $K=10^{-4}$ where the internal phases nodes are getting together in cascade as time increases as shown in Fig.\ref{casc}, where the number of independent states is decreasing with time. The dynamics completing the transition can be seen in movie 3 in supplemental material \textcolor{blue}{\cite{suppl}}. It is interesting to notice that the rotational motion of the phases shown in Fig.\ref{rexp}, remains all the way to synchronization, in spite of the cascade effect shown in Fig.\ref{casc}. This behavior is clearly seen in movie 3 of the Supplemental material.\\

\begin{figure}[h!]
\includegraphics[width=7.5cm, height=5cm]{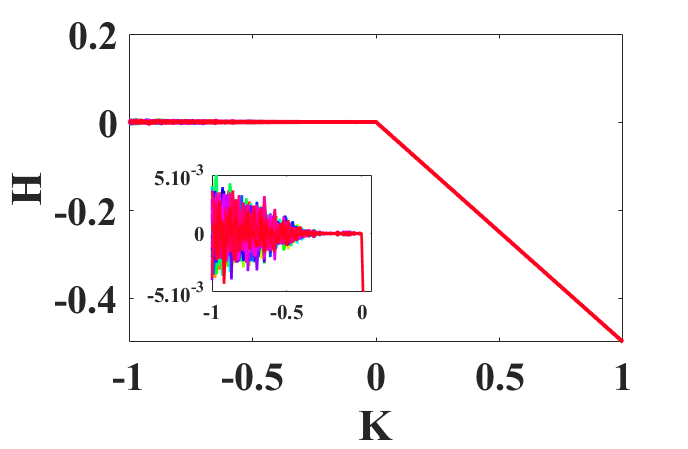}
\caption{Energy $H_i$ of each entity of the system as function of phase coupling $K$}. 
 \label{energy1}
\end{figure}

To continue the study of the evolution of the system towards the explosive transition to synchronization,  we explore the energy of the swarmalators systems for the case of constant frequencies which is plotted in Fig. \ref{energy1} as a function of $K$. We took advantage of previous works where we used the Hamiltonian formalism to understand the transition to synchronization  in a star network of coupled oscillators \textcolor{blue}{\cite{njougouo2020dynamics2}}, as well as to justify the existence of \textit{chimera and multi chimera states} \textcolor{blue}{\cite{simo2021traveling}}. \\

Near synchronization, when the spatial distance between the swarmalators is practically constant and since the coupling in Eq.\ref{ee2} is positive the individual phase dynamics is that of a conventional XY model and the Hamiltonian becomes \textcolor{blue}{\cite{hong2018active,kosterlitz1974XYmodel,kim2001xymodel,bhadra2018hamiltonianXYmodel,leoncini1998hamiltonianXYmodel,kosterlitz1973ordering}}:
\begin{equation}\label{Heq}
H_i =  - \frac{K}{{2N}}\sum\limits_{i \ne j}^N {\cos \left( {{\theta _i} - {\theta _j}} \right)}.
\end{equation}

However, the threshold value of the phase coupling $K$ where the synchronization appears is $K_c=0.005$ (and not zero as assumed before) and as shown on Fig. \ref{energy1}, the energy of the system decreases when the swarmalators phase synchronize. Since phase synchronization implies that $\theta_i=\theta_j$ (for all $i$ and $j$), Eq.\ref{Heq} becomes
\begin{equation}\label{Hcritiq}
H_c =  - \frac{{{K_c}}}{2}.
\end{equation}

From there, the minimum energy for which swarmalators synchronize is equal to $H_c$. This observation could explain that the apparently explosive transition to synchronization is in fact a process of energy loss where the elements synchronize to minimize the energy. Let us now take a look at the order of the phase transition when the system loses energy.\\

When the XY model represents the energy, Eq.\ref{ee2} reduces to the well known Kuramoto model \textcolor{blue}{\cite{kuramoto2002coexistence}}. Let us summarize how the present literature stands on this problem:\\

a) According to D. Pazó \textcolor{blue}{\citep{pazo2005thermodynamic}} the Kuramoto model has a first order first transition from incoherence to synchronization in the thermodynamic limit, provided the natural frequencies are evenly spaced or uniformly distributed in a finite range.\\

b) Gómez Gardeñes et al.\textcolor{blue}{\citep{gomez2011explosive}} have shown that, when the natural frequency and the degree of a node are equal, there is an explosive transition in a scale free network of Kuramoto oscillators. \\

c) Skardal et al.\textcolor{blue}{\cite{skardal2014disorder}} observed that adding infinitesimal disorder induces an explosive transition to synchronization for the case discussed in (b).\\

d) Leyva et al.\textcolor{blue}{\citep{leyva-2}} had extended results and although they claim that "a sharp, discontinuous phase transition is not restricted to the above rather limited and apparently opposite cases\textcolor{blue}{\citep{leyva-3,gomez2011explosive}}, but it constitutes, instead, a generic feature of the synchronization of networked
phase oscillators", their studies consider systems where the natural frequencies cannot be equal between themselves, or to the mean. This is not our case where we consider all natural frequencies equal to zero, nor that of Gómez Gardeñes et al, which has a very low probability of not repeating values of frequencies.\\

e) Hong and Martens \textcolor{blue}{\citep{hong2022first}} studied phase transitions in an XY model related to a variant of the Kuramoto model for coupled oscillators. They found that for the case without noise, the system shows features of a first-order phase transition at complete synchronization, while this transition is found to be continuous for the noisy case.\\

\begin{figure}[ht]
\includegraphics[width=4.25cm, height=3.5cm]{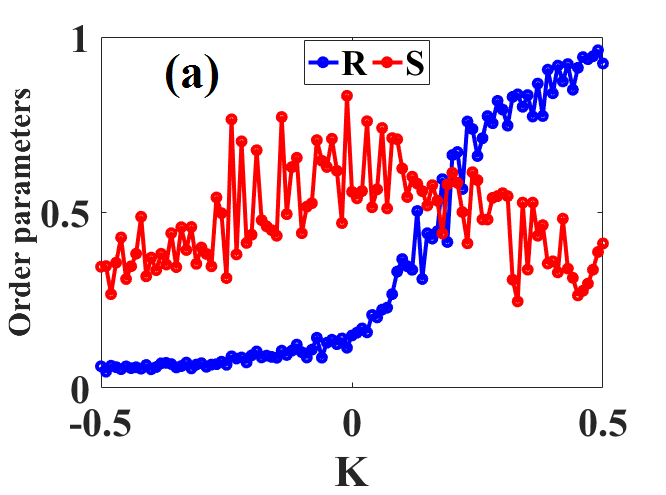}
\includegraphics[width=4.25cm, height=3.5cm]{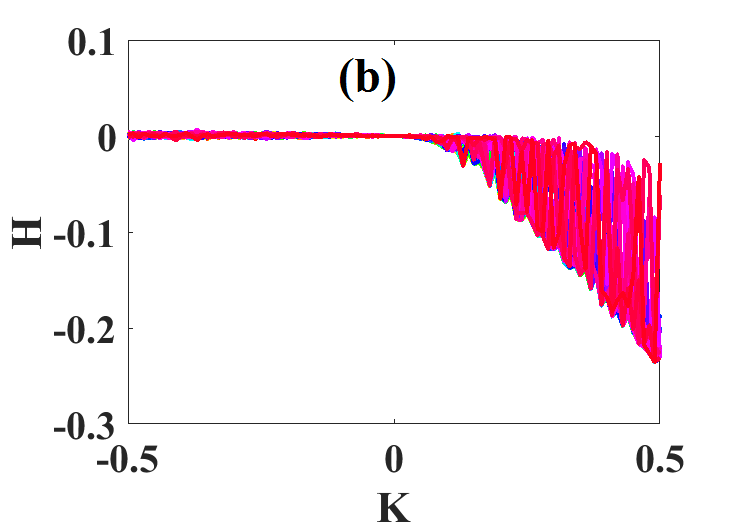}
\caption{Order parameter (a) and energy of swarmalators (b) as a function of phase coupling strength $K$ when the natural frequencies are randomly distributed (case a).}.
 \label{case2}
\end{figure}

 As shown by Eqs. \ref{ee1} and \ref{ee2}, our system belongs to the class discussed by Hong and Martens and it reduces to an XY model with a first order phase transition.
Meanwhile, the problem of different natural frequencies is not so clear. To search for an explanation we modify our system and study three cases where natural frequencies are distributed as follows:\\

a) The frequency is randomly distributed:

\quad Each oscillator is subject to a different frequency generated by a Rand function. From this random frequency distribution, we can see that although the system synchronizes, the transition is not explosive. See Fig. \ref{case2}.\\

b) The frequency is distributed in two groups of swarmalators:

\quad One group has a constant frequency while the other one is affected by noise, defined as follows $(w1=0.005)$:
\begin{itemize}
\item $1^{st}$ group:
$From\,\,1\,\,to\,\,\frac{N}{2}\,,\,\,\,\,\,w\left( i \right) = w_1$. \\ 
\item $2^{nd}$ group:
$From\,\,\,\frac{N}{2}+1\,\,to\,\,N\,,\,\,\,\,\,w\left( i \right) = w1\,\,+\varepsilon, \\ with\,\,\varepsilon  = {10^{ - 1}}$. \\
\end{itemize}

\begin{figure}[ht]
\includegraphics[width=4.25cm, height=3.5cm]{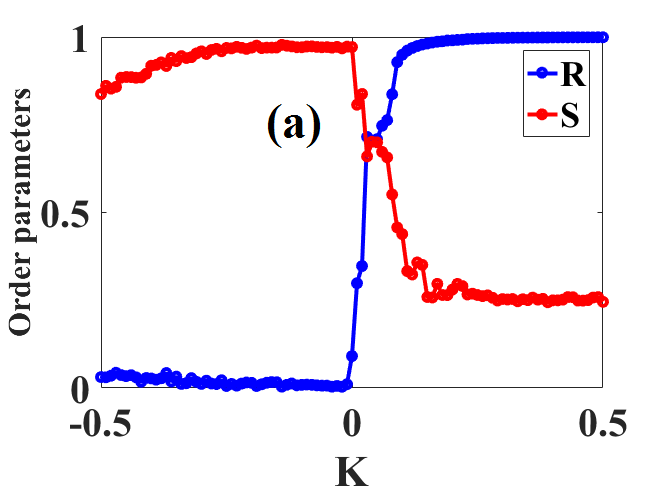}
\includegraphics[width=4.25cm, height=3.5cm]{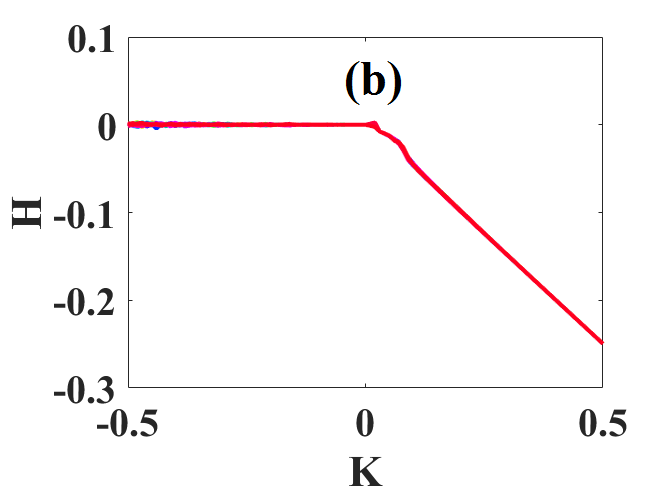}
\caption{Order parameter (a) and energy of swarmalators (b) as a function of phase coupling strength $K$ when the natural frequencies  are distributed in two groups (case b).}.
 \label{case3}
\end{figure}

Results are shown in Fig.\ref{case3}, where we notice that the sudden transition has disappeared.\\

\begin{figure}[ht]
\includegraphics[width=9cm, height=5cm]{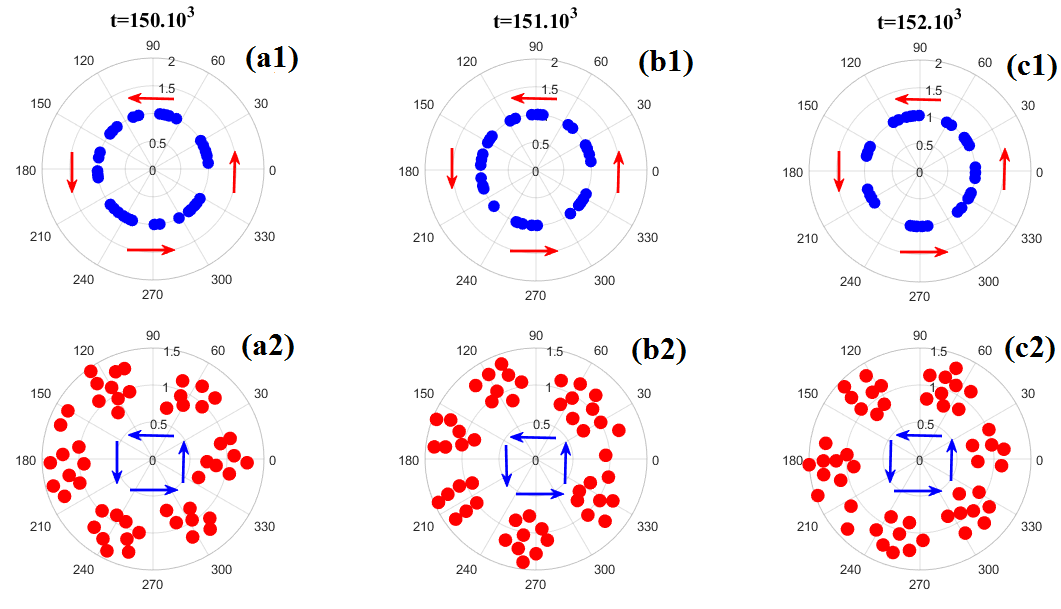}
\caption{Snapshots of all internal phases (a1,b1,c1) and nodes positions (a2,b2,c2) showing the changes in the positions of clusters for $t=150.10^3$, $t=151.10^3$ and $t=152.10^3$ when frequencies are distributed in two groups (case b).}.
 \label{2grp}
\end{figure}

However, for this distribution of frequencies and for $K\approx -0.015$ the swarmalators are regrouped in clusters, meaning $R \approx 0$, $S\approx 1$ and $V\ne 0$ properties which normally describe the Splintered Phase Wave state (SpPW) \cite{o2017oscillators}. In addition to the SpPW properties, these clusters are rotating as shown by snapshots in Fig.\ref{2grp}, where the internal phases of nodes are plotted in Fig.\ref{2grp}(a1, b1, c1) (blue dots) while their spatial positions, in polar coordinates, are given by Fig.\ref{2grp}(a2, b2, c2) (red dots) for three values of time $t=150.10^3$, $t=151.10^3$ and $t=152.10^3$. Thus this dynamics is a Rotating Splintered Phase Wave state (RSpPW) characterized by the value zero of the parameter $m_i = 0$ given by Eq.6 in \textcolor{blue}{\cite{mishra}} and defined by its expression $m_i = 1-0.5\left[\max(cos(\phi_i(t))) - \min(cos(\phi_i(t)))\right]$ for all nodes. Even if the total number of cluster in the considered case remains at seven, it appears that the number of nodes in the clusters can change with time. 

c) The frequency is distributed in three groups of swarmalators, distributed as follows:\\
\\ 
i) $1^{st}$ group: From $1$ to $15$, $w\left( i \right) = w1$.\\ \\
ii) $2^{nd}$ group: From $16$ to $31$, $w\left( i \right) = w1 + \varepsilon$ with $\varepsilon  = {10^{ - 2}}$ \\ \\
iii) $3^{rd}$ group: From $32$ to $N$, $w\left( i \right) = rand/10$.\\

We notice here that the phase transition towards synchronization is no longer explosive.

\begin{figure}[ht]
\includegraphics[width=4.25cm, height=3.5cm]{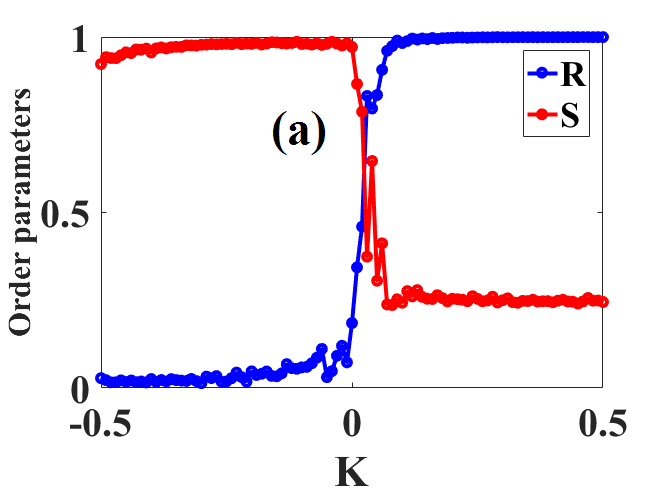}
\includegraphics[width=4.25cm, height=3.5cm]{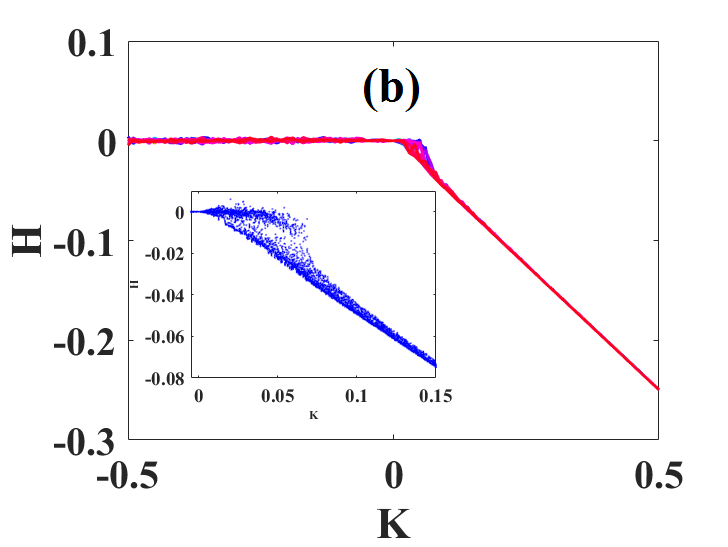}
\caption{Order parameter (a) and energy of swarmalators (b) as a function of phase coupling strength $K$ when the natural frequencies are distributed in three groups (case c).}.
 \label{case4}
\end{figure}

Therefore we studied in detail the transition to synchronization in this case and found that it is not a first order phase transition. First the oscillators cluster and then merge into asymptotic synchronization, as it can be seen in Fig. \ref{case4}(b) with emphasis in the inset.\\

Therefore, while we present here some of the possible cases of synchronization in swarmalators, a thorough study of this and related problems on phase transitions
in systems of mobile oscillators and their representation
by the Kuramoto model is necessary.\\

Summarizing, in this work, we have studied the behavior of systems where phase and spatial dynamics are coupled called swarmalators. The effect of positive and negative phase coupling strength on the dynamics of the system under the impact of initial conditions was shown. As proposed in \textcolor{blue}{\cite{o2017oscillators}}, the five steady states of swarmalators were characterized, and found that the system maps into an XY model with an explosive transition to the synchronization when the system is subject to an attractive $(K>0)$ and repulsive $(K<0)$ phase coupling. 
Based on the conventional mean field XY model, the mean energy of a system using the Hamiltonian formalism has been evaluated. This evaluation highlights that the transition of swarmalators to synchronize can be explained as a process of energy loss which leads them to synchronize when the critical value $H_c=-\frac{K_c}{2}$ reaches $K_c=0.005$, in a goal to minimize this energy. During this study, some expected results were obtained like the second order transition and pattern formation as described in \textcolor{blue}{\cite{o2017oscillators}}. However, in addition to the new state as the Rotational Splintered Phase Wave (RSpPW), another unexpected result here is the existence of first order transition without correlation between the natural frequencies of elements and the degree of their nodes. Indeed, This kind of transition, which was shown previously to happen when the natural frequencies of the oscillators are equal to the number of links it possesses \textcolor{blue}{\citep{gomez2011explosive}}, happens always that the natural frequencies of all oscillators are equal but it disappears when they show complete or partial disorder.  To better understand why the first-order transition occurs, we studied three main cases of natural frequency distribution. This hypothesis showed that how natural frequencies are distributed can considerably affect the type of transition in a swarmalator's systems. 
By making a detailed investigation of the dynamics before the transition we found that the internal phases undergo a rotational wave state, which we have not been able to explain why it appears there. More on this point will be done in future work.
\\

PL and HAC thank ICTP-SAIFR and FAPESP grant 2021/14335-0 for partial support. HAC thanks the University of Namur for hospitality.
TN thanks the University of Namur for financial support.
SJK, TN, PL and MoCLiS research group thank ICTP for the equipment, donated under the letter of donation Trieste $12^{th}$ August 2021.

\appendix

\section{Stability analysis of SWPW state}\label{Appx}

\begin{figure}[h!]
    \centering
 \includegraphics[width=9cm, height=6cm]{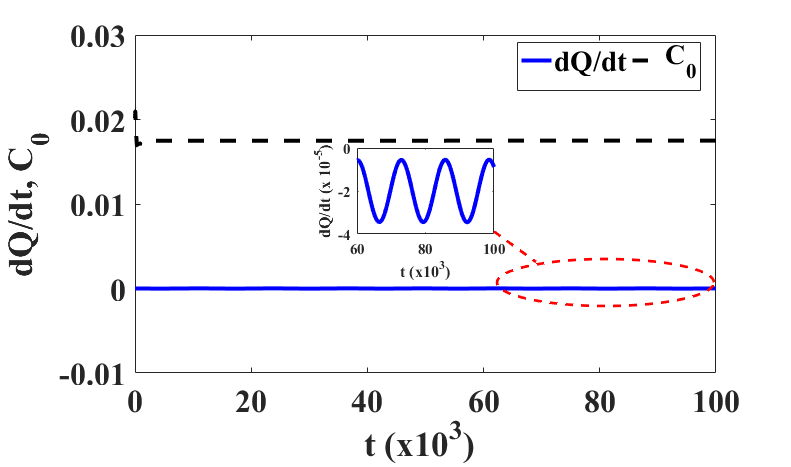}
 \caption{Stability condition of the SWPW state: time evolution of the Lyapunov derivative $\dot Q$ and the threshold constant $C_0$ for $K=0.00001$ and $J=1$.}
    \label{stabilitySWPW}
\end{figure}

Following the example of Seung-Yeal Ha et al. in \textcolor{blue}{\cite{ha2019emergent}}, we consider the following expressions: $$r(t) = X_2(t) - X_1(t)$$ and $$\varepsilon(t) = \theta_2(t) - \theta_1(t),$$ where $X_i$ and $\theta_i$ for $i=1,2$ are two randomly chosen elements of the network expressed by Eqs.3 and 4 with $v_i = w_i = 0$. After some manipulations developments and simplifications, the time derivatives of $r(t)$ and $\varepsilon$ give:\\
$\dot{r}(t) = -\left(A + J cos(\varepsilon(t)) - \displaystyle\frac{B}{|r(t)|}\right)\displaystyle\frac{r(t)}{|r(t)|}  $ and $\dot{\varepsilon}(t) = -\displaystyle\frac{2K}{|r(t)|} sin(\varepsilon (t)),$\\

which are the error state dynamics between both chosen elements. Since we are dealing with the SWPW state where the  swarmalators are static in space, $r(t) = r_{12}=r$  is a constant different from zero. 
As the spatial error cannot be zero the objective is to study the stability of our system around the point of equilibrium of the phases.
To do this, consider a small perturbation $\xi$ around the equilibrium trajectories ${\upsilon^*}$. The error of the system can be written as follows \\ $$\varepsilon  = \xi  + {\upsilon^*},$$ the new error system becomes $$\dot \xi  =  - \frac{{2K}}{r}\sin \left( {\xi  + {\upsilon^*}} \right).$$
From here, the goal is to show that, this system achieves at least practical stability in the sense of M. Sekieta and T. Kapitaniak \textcolor{blue}{\cite{sekieta1996practical}}.
Knowing that for the case of the Static Wing Phase Wave(SWPW), the derivative becomes \\ $$\dot \xi  = \frac{{2\left| K \right|}}{r}\left( {\sin \xi .\cos {\upsilon^*}\,\, + \,\,\sin {\upsilon^*}.\cos \xi } \right)$$
this lead to $$\dot \xi  = {C_1}\sin \xi \,\, + \,\,{C_2}\cos \xi ,$$ with ${C_1} = \frac{{2\left| K \right|}}{r}\cos {\upsilon^*}$ and ${C_2} = \frac{{2\left| K \right|}}{r}\sin {\upsilon^*}.$\\
From these expressions, to investigate the stability of this error system, let us choose the following Lyapunov function candidate:\\  $$Q = \frac{1}{2}{\sin ^2}\xi.$$
Its time derivative is given by $$\dot Q = \frac{1}{2}\left( {{C_1}\sin \xi \,\, + \,\,{C_2}\cos \xi } \right).sin2\xi, $$  By considering positive, the terms of the right-hand side of the derived of the Lyapunov candidate function, it imposes to choose 
$\dot Q$ such as 
$$\dot Q \le \frac{1}{2}\left| {{C_1}} \right|\left| {\sin \xi } \right|\left| {\sin 2\xi } \right|\ + \,\,\frac{1}{2}\left| {{C_2}} \right|\left| {\cos \xi } \right|\left| {\sin 2\xi } \right|$$
Knowing that $a.b \le \frac{{{a^2} + {b^2}}}{2}$ for $a > 0$, $b > 0$ \\
$$\dot Q \le \frac{{\left| {{C_1}} \right|}}{4}\left( {{{\sin }^2}\xi  + {{\sin }^2}2\xi } \right)\, + \,\,\frac{{\left| {{C_2}} \right|}}{4}\left( {{{\cos }^2}\xi  + {{\sin }^2}2\xi } \right),$$
Maximizing the sin and cosine functions, the Lyapunov derivative can be bounded as follows 
$$\begin{array}{l}
\dot Q \le \frac{1}{4}\left( {\left| {{C_1}} \right|\max \left( {{{\sin }^2}\xi } \right) + \left| {{C_2}} \right|\max \left( {{{\cos }^2}\xi } \right)} \right)\, + \,\\,\\
\,\,\,\,\,\,\,\,\,\,\,\,\,\,\,\,\,\,\,\,\,\,\,\,\,\,\,\,\,\left( {\frac{{\left| {{C_1}} \right|}}{4} + \frac{{\left| {{C_2}} \right|}}{4}} \right)\max \left( {{{\sin }^2}2\xi } \right)
\end{array}$$  Then we can write $$\dot Q \le \,\,\frac{1}{2}\left( {\left| {{C_1}} \right| + \left| {{C_2}} \right|} \right).$$ 
$$\dot Q \le \eta ,$$  with $$\eta  = \frac{{2\left| K \right|}}{r}>0.$$
Thus, it comes that,  the Static Wing Phase Wave State (SWPW) of swarmalators system described by Eqs.3 and 4 is stable if
$$\dot Q = \sum\limits_{i = 1}^N {\sum\limits_{i \ne j}^N {{{\dot Q}_{ij}}} },$$\\
which becomes, $${\dot Q} \le \sum\limits_{i = 1}^N {\sum\limits_{i \ne j}^N {\frac{{2\left| K \right|}}{{r}} }}$$ Therefore, it is established that the system defined by Eqs3 and 4 in the Static Wing Phase Wave state condition is practically stable \textcolor{blue}{\cite{kakmeni2010practical,louodop2017coherent}} since the time derivative of the Lyapunov function $\dot Q$ is bounded by a positive constant $\eta$, which proves the stability of the Static Wing Phase Wave. Following the previous study of Louodop et al. \textcolor{blue} {\cite{louodop2017coherent}}, this shows that, for any given initial condition, the distance between the trajectories remains constant and  according to M. Sekieta and T. Kapitaniak \textcolor{blue}{\cite{sekieta1996practical}} and R. Fermat and G. Solis - Perales \textcolor{blue}{\cite{fermatLyap2002}},  the error dynamics is not converging to zero but to a sufficiently small value that can be considered as the tolerance domain of the synchronization condition in the steady state, as the calculations prove. To verify this condition, we have plotted in Fig\ref{stabilitySWPW} the time evolution of the Lyapunov derivative $\dot Q$ and the threshold value $C_{0}=N(N-1)\times\eta$. This evolution adheres to the stability condition obtained earlier.

\nocite{*}

\end{document}